\documentclass[10pt,preprint]{aastex}
\usepackage{times}

\newcommand{\vect}[1]{\ensuremath{\mbox{\boldmath $#1$}}}

\slugcomment{to appear in the Astrophysical Journal Letters}
\shortauthors{AN \& GOULD}
\shorttitle{MICROLENS MASS MEASUREMENT}

\begin{document}

\title{Microlens Mass Measurement using Triple-Peak Events}

\author{Jin H. An and Andrew Gould}
\affil{Department of Astronomy, the Ohio State University,
140 W. 18th Ave., Columbus, OH 43210}
\email{jinhan,gould@astronomy.ohio-state.edu}

\begin{abstract}

We show that one can measure the effects of microlens parallax for
binary microlensing events with three well-measured peaks
-- two caustic crossings plus a cusp approach, and hence derive
the projected Einstein radius $\tilde r_{\rm E}$. Since 
the angular Einstein radius $\theta_{\rm E}$ is measurable from
finite source effects for almost any well-observed caustic crossing,
triple-peak events can yield the determination of the lens mass
$M = c^{-2}(4G)\tilde r_{\rm E} \theta_{\rm E}$.
We note that, to a certain extent,
rotation of the binary can mimic the effects of parallax, but it should
often be possible to disentangle parallax from rotation by making use of
the late-time light curve.
 
\end{abstract}

\section{Introduction}

When microlensing experiments were proposed \citep{Pa86},
it was clear that the lens masses would not be measured individually
on an event-by-event basis because, among the routine observables, only
the Einstein timescale $t_{\rm E}$ is related to the mass,
and only indirectly to it;
\begin{equation}
\label{eqn:tedef}
t_{\rm E} = \frac{\theta_{\rm E}}{\mu_{\rm rel}}
\,;\ \ \
\theta_{\rm E} \equiv \sqrt{\frac{2 R_{\rm Sch}}{D_{\rm rel}}}
\ .\end{equation}
Here
$D_{\rm rel} = \mbox{AU}/\pi_{\rm rel}$,
$\pi_{\rm rel}$ is the lens-source relative trigonometric parallax,
$\mu_{\rm rel}$ is the relative proper motion,
$\theta_{\rm E}$ is the angular Einstein radius,
and $R_{\rm Sch}$ is the Schwarzschild radius of the lens mass.
Rather, it would be necessary to interpret the results
statistically, to infer an average
mass scale of the lenses from the mean timescale of the observed
events together with statistical estimates of
$\pi_{\rm rel}$ and $\mu_{\rm rel}$.
Note that $D_{\rm rel}^{-1} = D_{\rm L}^{-1} - D_{\rm S}^{-1}$, 
where $D_{\rm L}$ and $D_{\rm S}$ are
the distances to the lens and the source.

\citet{Go92} showed that it was possible to
extract the individual lens mass $M$ 
if one could measure two ``observables'':
the angular Einstein radius $\theta_{\rm E}$,
and the projected Einstein radius $\tilde r_{\rm E}$;
\begin{equation}
\tilde r_{\rm E} \equiv \theta_{\rm E} D_{\rm rel} =
\sqrt{2 R_{\rm Sch} D_{\rm rel}}
\,,\end{equation}
\begin{equation}
\label{eqn:retildedef}
M
= \frac{c^2}{4 G} \tilde r_{\rm E} \theta_{\rm E}
\ .\end{equation}
The microlens parallax $\vect{\pi}_{\rm E}$
re-expresses $\tilde r_{\rm E}$ in a convenient way. It is a two-dimensional
vector whose magnitude is given by
\begin{equation}
\label{eqn:micropar}
|\vect{\pi}_{\rm E}| \equiv \frac{\pi_{\rm rel}}{\theta_{\rm E}} =
\frac{\mbox{AU}}{\tilde r_{\rm E}}
\,,\end{equation}
and whose direction is that of the lens-source relative proper motion.
Measurement of $\theta_{\rm E}$ requires that the Einstein ring be compared
with some ``angular ruler'' on the plane of the sky, while to measure
$\tilde r_{\rm E}$, one must compare the Einstein ring with some ``physical
ruler'' in the observer plane.

The only angular ruler to be used for this purpose to date is the angular
radius of the source $\theta_\ast$
which can be estimated from its de-reddened color and magnitude,
and the empirical color/surface-brightness relation
\citep[eg.][]{vB99}.
When the source passes very close to or directly over a caustic
(zone of formally infinite magnification) in the lens magnification pattern,
the finite size of the source affects the microlensing light curve.
By analyzing this finite-source affected light curve, one can measure
$\rho_\ast$ $\equiv \theta_\ast/\theta_{\rm E}$, and thus
$\theta_{\rm E}$ as well.
While this idea was originally proposed for point-mass lenses,
which have point-like caustics \citep{Go94, NW94, WM94},
in practice it has been mainly used for binary lenses,
which have line-like caustics and hence much larger cross sections
\citep{MA02, MA03, PL01, PL02, PL03, SMC}.

Typically $\tilde r_{\rm E}$ is of order several AU, and so the
natural scale of the physical ruler must be $\sim 1\ \mbox{AU}$.
This is the case for the two main methods proposed to measure
$\tilde r_{\rm E}$. First, a satellite in solar orbit would see a substantially
different event from the one observed on Earth, and from the difference,
one can determine $\tilde r_{\rm E}$ \citep{Re66, Go95}. Second, the reflex of
the Earth's orbital motion induces a wobbling of
the source's passage through the Einstein ring,
which in turn perturbs the light curve, and $\tilde r_{\rm E}$
can be determined from this perturbation \citep{Go92}. To date, this
latter method is the only one by which $\tilde r_{\rm E}$ has been measured
\citep*{MA01, Ma99, OGLE, MOA, OGLEII, O9932, Be01}.
Unfortunately, this method requires that the event be rather long,
$t_{\rm E} \ga 90\ \mbox{days}$. To the extent that the Earth's
motion can be approximated as rectilinear, it is not possible
even in principle to detect this effect.
Over the relatively short timescales of typical
events $t_{\rm E} \sim 20\ \mbox{days}$, on the other hand
the Earth's motion can be approximated
as uniform acceleration which can lead to a potentially detectable
asymmetry in the light curve \citep*{GMB94}. However, this would yield
a measurement of only the projection of $\vect{\pi}_{\rm E}$,
along the direction of the Earth's acceleration
vector at the peak of the event. Hence, even if this asymmetry were
detected, it would give only an upper limit on $\tilde r_{\rm E}$.
To measure both components of $\vect{\pi}_{\rm E}$ 
requires that the event last long
enough for the Earth's acceleration vector to change substantially
while the source lies within the Einstein ring.

However, it is possible to make parallax
measurements using much shorter baselines. During a caustic crossing
of a binary lens, the flux can change by several magnitudes as the
caustic transits the star. The source radius projected
onto the observer plane $\theta_\ast D_{\rm rel}$ 
can be of order $\sim 100\ \mbox{R}_\earth$ for a main-sequence star.
Hence, observers on
two continents could see fluxes differing by a few percent.
This would again yield only one projection of $\vect{\pi}_{\rm E}$,
but by repeating this procedure for the two caustic crossings of a
single event, or by observing one crossing from three continents, one
could obtain a full measurement of parallax \citep{HW95, GA99}. An important
feature of this approach is that it also allows to measure
$\theta_{\rm E}$, which can be obtained from any caustic crossing.
Hence, if this method were ever applied, it would yield the mass through
equation~(\ref{eqn:retildedef}). Unfortunately, the method requires
extraordinary cooperation from the event itself, from the observatory
directors, and from the weather. Consequently, it has not to date been
successfully carried out.

It is expected that
the use of the high-precision interferometric astrometry, most
notably the \emph{Space Interferometry Mission} (SIM) will be
the best way to measure the individual lens mass \citep*{Pa98,BSV98}.
With 10-$\mu$as level astrometry of SIM, one can directly measure
$\pi_{\rm rel}$. Furthermore, by monitoring the displacements
in the light centroid caused by microlensing, $\theta_{\rm E}$ can be also
directly measured \citep*{HNP95, MY95, Wa95}. Unlike most techniques
mentioned earlier, neither of these measurements requires
the event to be of a special class such as long
time-scale or caustic-crossing events. That is, individual
lens masses can be measured for common point-source/point-lens events.
 
Here, we propose a new method to measure parallax for a subclass
of caustic crossing binaries. The method requires the event to have
at least three separate well-defined photometric peaks so that the
source position can be precisely located relative to the lens geometry
at three distinct times. Two of the peaks should be
caused by caustic crossings: an entrance and an exit.
It would be best for the third peak to
be a cuspy-caustic crossing, but the chance of two
fold-caustic crossing accompanied by an additional cusp crossing
is low. In the following, we investigate events where the
third peak is due either to a cusp crossing or to
a generic cusp approach.

\section{Overview}

If the event has two well-observed caustic
crossings, plus a well-observed ``cusp approach'', these features
then provide nine ``empirical parameters''
-- parameters that can be measured directly from the light curve,
without appealing to a model that depends upon a global lens geometry.
They are the maximum flux $F_{\max}$, the half-duration $\Delta t$, and 
the time at the maximum $t_{\max}$ of each of the three peaks.
The uncertainties in the measurements of these quantities are approximately
\begin{equation}
\label{eqn:sigmaeq}
\frac{\sigma(F_{\max})}{F_{\max}} \sim
\frac{\sigma(t_{\max})}{\Delta t} \sim
\frac{\sigma(\Delta t)}{\Delta t} \sim 
\frac{\sigma_{\rm ph}}{\sqrt N} = 
10^{-3} \left(\frac{N}{100}\right)^{-1/2} \frac{\sigma_{\rm ph}}{1\%}
\,,\end{equation}
where $\sigma_{\rm ph}$ is the mean fractional error of each of the $N$
photometric measurements taken over the bump.  Since $\Delta t$ is generally
of order a day, the errors in $\Delta t$ and $t_{\max}$ measurements
can easily be of
order a minute or smaller. The fact that six independent times can be measured
with such precision, typically four or five orders of magnitude smaller than the
characteristic timescale of the event $t_{\rm E}$, is what makes the parallax
measurement feasible.

In principle, any localizable feature found in the light curve may
provide similar empirical parameters. What then makes peaks more significant
as information posts than other features such as minima
or points of inflection? The answer to this question is
that the high precision measurements attainable for peaks
can rarely be achieved for empirical parameters
associated with the other localizable features,
so that their merit as signature beacons is much weaker than for peaks.
For instance, minima in microlensing light curves typically have
widths that are ten times greater than peaks do (and by definition
less flux as well). From equation~(\ref{eqn:sigmaeq}),
$\sigma(t)\propto\sqrt{\Delta t/F}$, so that it is substantially more
difficult to localize minima than peaks. Here, we assume Poisson
noise ($\sigma_{\rm ph} = \sqrt{F}/F)$, and the same sampling frequency
(i.e.\ $N\propto\Delta t$). Points of inflection are even more
difficult to localize than minima.

Ordinarily,
under the approximation that the lens-source relative motion is
rectilinear, to specify a binary-lens light curve requires seven
global (or geometric) parameters:
$d$, the binary separation in units of $\theta_{\rm E}$;
$q$, the binary mass ratio;
$\alpha$, the angle of the source-lens relative motion
relative to the binary axis;
$t_{\rm E}$, the Einstein timescale;
$u_0$, the minimum angular separation
between the source and the binary center in units of $\theta_{\rm E}$;
$t_0$, the time at this minimum;
$\rho_\ast$, the source size in units of $\theta_{\rm E}$.
(In addition, to transform the magnification curve to the specific photometric
system of the observations, one also needs
limb-darkening parameters for each wave band of observations;
plus the source flux $F_{\rm s}$ and background flux $F_{\rm b}$,
for each telescope and wave band.)
However, since the actual motion is not rectilinear, these
seven parameters will not be adequate to describe the event
for very high precisions, and
in particular, subtle inconsistencies will be introduced among the
nine precisely measured quantities mentioned earlier. 
We now show how these inconsistencies
can lead to a parallax measurement.

\section{Measurement of Parallax}

In practice, the parallax will be measured by
multi-dimensional fitting and subsequent $\chi^2$ minimization.
However, it is instructive for two reasons to identify 
in a systematic way the features
of the event that permit $\vect{\pi}_{\rm E}$ to be measured.
First, this enables one to predict
when an event will have a measurable $\vect{\pi}_{\rm E}$.
Second, there are technical difficulties associated with $\chi^2$ minimization,
and these can be ameliorated if the model parameterization is modified
to reflect the underlying physics \citep{PLCS}.

To understand how $\vect{\pi}_{\rm E}$ is measured, 
we first show how some
of the nine empirical parameters are related to
one another in the absence of parallax, i.e., $\vect{\pi}_{\rm E}=0$.  
We will initially assume that the lens geometry ($d,q$) and the
fluxes $F_{\rm s}$ and $F_{\rm b}$ are known a priori.
Then, at the peak of the cusp approach $t_{\max,{\rm ca}}$,
the source position within the Einstein ring
$\vect{u}_{\rm ca}$ can be
determined very precisely from the precise measurements of 
$t_{\max,{\rm ca}}$ and $F_{\max,{\rm ca}}$.
That is, at this time $t_{\max,{\rm ca}}$,
the source must be somewhere along the cusp-approach
ridge line and its position on that line is determined from the 
inferred magnification $A = (F_{\max,{\rm ca}}-F_{\rm b})/F_{\rm s}$.  

Now consider the angles at which
the source crosses each caustic line, $\phi_1$ and $\phi_2$.
The half-duration of a caustic crossing is given by 
$\Delta t_{\rm cc} = t_{\rm E} \rho_\ast \csc\phi$.
If this motion is assumed to be rectilinear,
then $t_{\rm E}$ becomes a well-defined quantity over the whole duration of
the event, and thus,
\begin{equation}
\label{eqn:phicomp}
\frac{\csc\phi_2}{\csc\phi_1} = 
\frac{\Delta t_{{\rm cc},2}}{\Delta t_{{\rm cc},1}}
\ .\end{equation}
The source trajectory is then the straight line 
that passes through $\vect{u}_{\rm ca}$
and has caustic-crossing angles that satisfy equation~(\ref{eqn:phicomp}).
From that equation, this angle can be determined with a precision of
\begin{equation}
\label{eqn:alphaerr}
\sigma(\alpha) = \frac{1}{|\cot\phi_2 - \cot\phi_1|}
\left\{ 
\left[\frac{\sigma(\Delta t_{{\rm cc},2})}{\Delta t_{{\rm cc},2}}\right]^2 +
\left[\frac{\sigma(\Delta t_{{\rm cc},1})}{\Delta t_{{\rm cc},1}}\right]^2
\right\}^{1/2}
\,,\end{equation}
which can be estimated using equation~(\ref{eqn:sigmaeq}).

Next, we turn to the times of the two caustic 
crossings $t_{0,{\rm cc},1}$ and $t_{0,{\rm cc},2}$,
at which the center of the source lies on the caustic lines.
These are not the same as the times of peak flux
$t_{\max,{\rm cc},1}$ and $t_{\max,{\rm cc},2}$,
but they can be determined to the same precision 
(see eq.~[\ref{eqn:sigmaeq}]) and are more convenient to work with.
Let $\vect{u}_{{\rm cc},1}$ and $\vect{u}_{{\rm cc},2}$ be the positions in the
Einstein ring of the two caustic crossings.  For rectilinear motion, these
satisfy the vector equation,
\begin{equation}
\label{eqn:rectilinear}
\frac{\vect{u}_{{\rm cc},2}-\vect{u}_{{\rm cc},1}}
{t_{0,{\rm cc},2}-t_{0,{\rm cc},1}} =
\frac{\vect{u}_{\rm ca}-\vect{u}_{{\rm cc},1}}
{t_{\max,{\rm ca}}-t_{0,{\rm cc},1}}
\ \left(= \left|\vect{\mu}_{\rm E}\right|\right)
\ .\end{equation}
Here, 
$|\vect{\mu}_{\rm E}|=t_{\rm E}^{-1}$.
However, if the acceleration due to the Earth's orbital motion
is not negligible, equation~(\ref{eqn:rectilinear})
will not in general be satisfied. This can be quantified
in terms of $\delta t_2$, the difference between the measured value of
$t_{0,{\rm cc},2}$ and the one that would be predicted from the other measured
parameters on the basis of equation~(\ref{eqn:rectilinear}),
\begin{equation}
\label{eqn:ttwodif}
\delta t_2 \equiv (t_{0,{\rm cc},2} - t_{0,{\rm cc},1}) -
\frac{|\vect{u}_{{\rm cc},2}-\vect{u}_{{\rm cc},1}|}
{|\vect{u}_{\rm ca}-\vect{u}_{{\rm cc},1}|}
(t_{\max,{\rm ca}} - t_{0,{\rm cc},1})
\ .\end{equation}
To evaluate the relation between $\delta t_2$ 
and $\vect{\pi}_{\rm E}$, we make the
approximation that the Earth's acceleration vector projected on the
sky $\vect{a}_{\earth,\perp}$ is constant for the duration 
of the caustic crossings and
cusp approach.  We first note that the magnitude of this acceleration is
related to the parallax by
\begin{equation}
\label{eqn:aoverr}
\frac{|\vect{a}_{\earth,\perp}|}{\tilde r_{\rm E}} =
|\vect{\pi}_{\rm E} \sin\psi| 
\left(\frac{\Omega_\earth}{r_\earth/\mbox{AU}}\right)^2
\,,\end{equation}
where
$\psi$ is the angle between the lines of sights towards the Sun 
and the event from the Earth
$r_\earth$ is the distance between the Sun and the Earth,
during the caustic crossings, and $\Omega_\earth = 2 \pi\ \mbox{yr}^{-1}$.
Then, after some algebra, we find
\begin{equation}
\label{eqn:deltat2}
\frac{\delta t_2}{t_3-t_2} = \frac{|\sin\psi| t_{\rm E}}{2}
\left( \frac{\Omega_\earth}{r_\earth/\mbox{AU}} \right)^2
\{ -\pi_{{\rm E},\parallel} (t_2 - t_1) + 
\pi_{{\rm E},\perp} [(t_3-t_2) \cot\phi_2 - (t_3-t_1) \cot\phi_1] \}
\,,\end{equation}
where ($\pi_{{\rm E},\parallel},\pi_{{\rm E},\perp}$) are the components of
$\vect{\pi}_{\rm E}$ parallel and perpendicular to $\vect{a}_{\earth,\perp}$,
and where we have used the simplified notations, 
$t_{0,{\rm cc},i}\rightarrow t_i$ and $t_{\max,{\rm ca}}\rightarrow t_3$.
Hence, by measuring $\delta t_2$ one can determine a particular projection
of $\vect{\pi}_{\rm E}$ 
whose components are given by equation~(\ref{eqn:deltat2}).

However, since $\vect{\pi}_{\rm E}$ is a two-dimensional vector, 
measurements of two independent components are required for its
complete determination. A second constraint is available from the width
of the cusp approach $\Delta t_{\rm ca}$. We define
$\delta \ln \Delta t_{\rm ca}$ in analogy to $\delta t_2$
and after some more algebra we find,
$$
\delta\ln\Delta t_{\rm ca} =
\frac{|\sin\psi| t_{\rm E}}{\cot\phi_2 - \cot\phi_1}
\left( \frac{\Omega_\earth}{r_\earth/\mbox{AU}} \right)^2
\{ \pi_{{\rm E},\parallel}
[(t_3-t_2) \cot\phi_1 + (t_2-t_1)  \cot\phi_3 - (t_3-t_1) \cot\phi_2]
\\$$\begin{equation}
+ \pi_{{\rm E},\perp}
[ (t_3-t_1) \cot\phi_3 \cot\phi_1 - (t_3-t_2) \cot\phi_3 \cot\phi_2
- (t_2-t_1) \cot\phi_2 \cot\phi_1 ] \}
\ .\label{eqn:deltalndelta}
\end{equation}
Measurement of $\delta\ln\Delta t_{\rm ca}$ therefore gives
another projection of $\vect{\pi}_{\rm E}$.
Furthermore, it is possible to obtain a third constraint on 
$\vect{\pi}_{\rm E}$ from the behavior of the light curve
after the source has left the caustic but
while it still remains within the Einstein ring.
For rectilinear motion,
the Einstein time scale is well determined 
(cf.\ eq.~[\ref{eqn:rectilinear}]). If the late-time
light curve drops off faster or slower than indicated by this timescale,
it implies that the Earth's acceleration vector has a component aligned
with the direction of lens-source relative motion. The effect is similar
to the one identified by \citet{GMB94} but can be measured more easily
because both $t_{\rm E}$ and $t_0$ are determined very precisely from the
light curve around the caustic-crossing region.
The relative orientation of these three constraints depends on the
details of the lens geometry.  In principle, they could all be roughly 
parallel, but this is unlikely; in general, it should be possible to
combine the three projections to measure both components of
$\vect{\pi}_{\rm E}$.
If the event is sufficiently long (typically $t_{\rm E} \ga 60\ \mbox{days}$),
so that the Earth moves $\ga 1\ \mbox{radian}$ during an Einstein timescale
then the late time behavior of the light curve will give information about
both components of $\vect{\pi}_{\rm E}$.  In this case there would be the
fourth constraint.

The exact expression for the errors in the components of $\vect{\pi}_{\rm E}$ 
obtained from the first two constraints can be derived from
equations~(\ref{eqn:deltat2}) and (\ref{eqn:deltalndelta}), but these are
extremely complicated and for that reason not very interesting.
However, reasonable estimates of these errors can be made as follows.
First, we note that the error from applying the $\delta t_2$ constraint
is dominated by the problem of determining the change in $\alpha$ from
the parallax to the non-parallax case. The error in the measurement of
$\alpha$ is given by equation~(\ref{eqn:alphaerr}). On the other hand, the
change in $\alpha$ (chosen for definiteness to be the angle at the time
of the cusp approach) is given by
\begin{equation}
\label{eqn:deltaalpha}
(\cot\phi_2 - \cot\phi_1) \delta\alpha =
\frac{2 \delta t_2}{t_3 - t_2}
\ .\end{equation}
Hence the fractional error in $\pi_{{\rm E},\delta t_2}$, 
the projection of $\vect{\pi}_{\rm E}$ measured by this constraint,
is of the order of
\begin{equation}
\label{eqn:errest1}
\frac{\sigma(\pi_{{\rm E},\delta t_2})}{\pi_{{\rm E},\delta t_2}} \sim
\frac{\sigma(\alpha)}{\delta \alpha} \sim
\frac{N^{-1/2}\,\sigma_{\rm ph}}
{|\sin\psi| \Omega_\earth^2 (t_2-t_1) t_{\rm E} \pi_{\rm E}} \,,
\end{equation}
where we have made use of equations~(\ref{eqn:sigmaeq}) and
(\ref{eqn:alphaerr}). For typical bulge parameters,
$|\sin\psi|\sim 0.2$ and $\pi_{\rm E} \sim 0.1$, and with
good photometric coverage of the caustic crossings,
$N^{-1/2}\sigma_{\rm ph} \sim 10^{-3}$, the fractional error is 
$\sim 0.17 
\times
(t_{\rm E}/50\ \mbox{days})^{-1} 
\times
[(t_2-t_1)/20\ \mbox{days}]^{-1}$
and hence $\pi_{{\rm E},\delta t_2}$ should plausibly be measurable.
For the other constraint one finds similar expressions.

We now relax our assumption that 
$d$, $q$, $F_{\rm s}$, and $F_{\rm b}$ are known a priori. Actually, 
if $d$ and $q$ are known, $F_{\rm s}$ and $F_{\rm b}$ can
be easily determined from, for example, 
the baseline flux $F_{\rm s} + F_{\rm b}$, and
the minimum flux observed inside the caustic $A_{\min} F_{\rm s} + F_{\rm b}$.
Here $A_{\min}$ is the magnification at minimum which, for fixed ($d,q$), is
virtually independent of the minor adjustments to the trajectory due to
parallax. However, it is still necessary to relax the assumption that $d$
and $q$ are known. In fact, these must be determined simultaneously with the
parallax because changes in ($d,q$) can have effects on the relative times
of the caustic and cusp crossing and on the crossing angles, just as parallax
can. Nevertheless, there are numerous other constraints on ($d,q$) coming
from the overall light curve, and so while ($d,q$) and $\vect{\pi}_{\rm E}$
can be expected to be correlated, they should not be completely degenerate.
Hence, even allowing for degradation of the signal-to-noise ratio
for the $\vect{\pi}_{\rm E}$ measurement due to the correlations
between the projections of $\vect{\pi}_{\rm E}$ and ($d,q$),
it should be possible to measure $\vect{\pi}_{\rm E}$
with reasonable precision for events with two 
well-covered caustic crossings and a well-covered cusp approach.

\section{Effect of Binary Rotation}

The measurement of the parallax discussed in the previous sections
essentially relies upon the failure of the rectilinear approximation
of the lens-source relative motion when the trajectory is
over-constrained by available observations. For a relatively short time scale,
what is actually measured from this is an instantaneous acceleration on
the source trajectory and this acceleration may contain
significant contributions by other effects,
a notable example of which is the binary rotation around its center of mass.
The rotational motion of the binary lens projected onto the plane of the sky 
is observable in terms of a contraction or expansion of the binary separation
$\dot d$, and the lateral rotation of the binary axis with respect to
the fixed direction on the sky $\omega$. The apparent observable result
of the latter effect may heuristically be understood as
a centripetal acceleration $\sim u_{\rm c} \omega^{2}$
on the source motion relative to the (static) lens system,
where $u_{\rm c}$ is the angular extent of the caustic in units of the Einstein
ring.
For a face-on circular binary orbit observed at the ecliptic pole,
the ratio of the Earth's acceleration to the projected acceleration due to
the binary orbit is then
\begin{equation}
\label{eqn:rotation}
\frac{a_\earth}{\tilde a_{\rm bin}} = 
\frac{\pi_{\rm E}}{u_{\rm c}} \left(\frac{P}{\mbox{yr}}\right)^2
\ .\end{equation}
Thus, depending on these parameters, either parallax or
binary rotation could dominate.  In addition, for the general case,
both the Earth and binary acceleration would be reduced by possibly
very different projection factors (while the projection factor
for the binary rotation is from the orbital inclination of the binary,
the parallactic projection is due to the angle between the direction
to the ecliptic pole and the line of sight to the event).
Although the effect of $\dot d$ is
in general not expressible analytically, one can expect
it to be of similar order of magnitude to the one of $\omega$.

Unambiguous determination of the parallax therefore requires
the measurement of the acceleration at least at two
different times, or not less than four independent constraints on
the projection of the acceleration at different times. We
argued above that there are generically at least three constraints for the
types of events under discussion and that there is a fourth constraint for
sufficiently long events.  In this latter case, parallax can be unambiguously
discriminated from rotation.
But even when the event is short, so that there are only
three constraints, it may still be feasible to measure the parallax.
We first note that the rotation cannot significantly affect the late-time 
light curve -- i.e., $t_{\rm E}$ is not influenced by rotation --
for it does not affect separations between the source and the binary
center of masses. Hence, if one component of  $\vect{\pi}_{\rm E}$ is
measured from the late-time light curve, it may be possible using
equation (\ref{eqn:rotation}) and Kepler's Third Law, to show that parallax is
more important than rotation, in which case the caustic crossings can
be used to determine the full parallax $\vect{\pi}_{\rm E}$.

In brief, there is good reason to hope that events with three well-defined
bumps can yield parallax measurements, and hence mass measurements.
Whether this will be possible for any particular such event can only
be determined by detailed modeling.

\acknowledgements
{\center \bf ACKNOWLEDGEMENTS}

This work is supported in part by the National Science Foundation (NSF)
grant AST 97-27520 and in part by the Jet Propulsion Laboratory (JPL)
contract 1226901. Work by JA is supported by the Presidential
Fellowship from the Graduate School of the Ohio State University.

\clearpage

\end{document}